# Reconfigurable microwave photonic Fano filters based on optical Kerr microcombs

*Qi Zou, Jiayang Wu, Senior Member, IEEE, Yang Sun, Yang Li, Guanghui Ren, Senior Member, IEEE, Thach G. Nguyen, Xingyuan Xu, Bill Corcoran, Sai T. Chu, Roberto Morandotti, Fellow, IEEE, Fellow, Optica, Arnan Mitchell, Fellow, Optica, and David J. Moss, Life Fellow, IEEE, Life Fellow, Optica*

***Abstract*— Microwave photonic (MWP) Fano filters, featuring asymmetric filter shapes that enable steep spectral transitions, are attractive for high-bandwidth microwave signal processing such as frequency discrimination. However, achieving both steep spectral transitions and a high degree of reconfigurability remains challenging for conventional methods relying on direct mapping of Fano resonances generated by optical filters. Here, we propose and experimentally demonstrate a new way for realizing MWP Fano filters based on a microcomb-driven transversal filter system. Leveraging the large number of comb lines provided by microcombs as discrete taps, the transversal filter system can synthesize filter response that closely resembles Fano resonances, yielding high roll-off rates and slope rates up to ~33.8 dB / GHz and ~25.7 dB / GHz in our experiments, respectively. In addition, by simply programming the tap coefficients without changing any hardware, highly reconfigurable filter response can be realized. We experimentally demonstrate independent tuning of all three Fano characteristic parameters, including the asymmetry factor, resonance linewidth, and center frequency. These results verify the effectiveness of our approach for implementing highly reconfigurable MWP Fano filters with steep spectral transitions, offering strong versatility for meeting diverse requirements in practical applications.**

## I. INTRODUCTION

Microwave photonic (MWP) filters, which perform filtering functions in the microwave (MW) frequency band using photonic technologies, offer attractive advantages of broad operation bandwidths that overcomes the intrinsic bandwidth bottleneck of electronic devices, low loss that improves signal-to-noise ratios, and strong immunity to electromagnetic interference [1-4]. Nowadays, MWP filters have found wide applications for high-bandwidth MW signal processing in modern communication, radar, and sensing systems [5-8]. As a fundamental class of MWP filters, MWP Fano filters, featuring asymmetric filter shapes that enable steep spectral transitions, are particularly attractive for high-sensitivity frequency discrimination and instantaneous frequency measurement (IFM) [9-11].

Conventional MWP Fano filters are typically realized by using a single optical carrier to map the response of Fano resonances generated by optical resonators into the MW domain [9-11]. In these filters, Fano resonances are generated by engineering the interference between a discrete localized state and a continuum state [12-14], and high quality (Q) factors of the optical resonators are needed to enable steep spectral transitions, thereby enhancing the sensitivity for frequency discrimination. However, the high Q factors also lead to increased sensitivity to resonance drift and fabrication tolerances, thus imposing stringent requirements on resonance alignment and thermal stabilization for long-term operation [15]. In addition, these filters usually lack reconfigurability, although minor adjustments can be made by introducing PN junctions [16, 17] or thermo-optic heaters [18, 19]. In particular, the characteristic parameters of Fano filters, including the asymmetry factor, resonance linewidth, and center frequency, are determined by multiple structural parameters of the optical resonators, making it challenging to tune one parameter independently without affecting the others. The above limitations collectively make it challenging to realize MWP Fano filters that simultaneously offer steep

This work was supported in part by the Australian Research Council (ARC) Centre of Excellence Project in Optical Microcombs for Breakthrough Science (COMBS) under Grant CE230100006, and in part by the ARC Discovery Projects under Grant DP150104327, Grant DP190102773, and Grant DP190101576. (*Corresponding authors: Jiayang Wu; David J. Moss.*)

Qi Zou, Jiayang Wu, Yang Sun, Yang Li, and David J. Moss are with the Optical Sciences Center, Swinburne University of Technology, Hawthorn, VIC 3122, Australia, and also with the ARC Centre of Excellence in Optical Microcombs for Breakthrough Science (COMBS), (e-mail: nzou@swin.edu.au; jiayangwu@swin.edu.au; yangsun@swin.edu.au; yangli@swin.edu.au; dmoss@swin.edu.au).

Guanghui Ren, Thach G. Nguyen, and Arnan Mitchell are with Integrated Photonics and Applications Centre, School of Engineering, RMIT University, Melbourne, 3000 VIC, Australia, and also with the ARC Centre of Excellence in Optical Microcombs for Breakthrough Science (COMBS), (e-mail: guanghui.ren@rmit.edu.au, thach.nguyen@rmit.edu.au; arnan.mitchell@rmit.edu.au).

Xingxuan Xu is with State Key Laboratory of Information Photonics and Optical Communications, Beijing University of Posts and Telecommunications, Beijing 100876, China (e-mail: xingyuanxu@bupt.edu.cn).

Bill Corcoran is with Photonic Communications Laboratory, Dept. Electrical and Computer Systems Engineering, Monash University, Clayton, VIC, Australia, and also with the ARC Centre of Excellence in Optical Microcombs for Breakthrough Science (COMBS), (e-mail: bill.corcoran@monash.edu).

S. T. Chu is with the Department of Physics, City University of Hong Kong, Hong Kong (e-mail: saitchu@cityu.edu.hk).

Roberto Morandotti is with the INRS – Énergie, Matériaux et Télécommunications, Varennes, QC J3X 1S2, Canada (e-mail: morandotti@emt.inrs.ca).

spectral transitions and a high degree of reconfigurability.

In this work, we propose and experimentally demonstrate MWP Fano filters based on a microcomb-driven transversal filter system, enabling both steep spectral transitions and a high degree of reconfigurability. By using the large number of comb lines provided by optical microcombs as discrete taps, the transversal filter system can synthesize filter response that closely resembles Fano resonances through coherent combination of delayed and weighted taps. Experimental results show that high roll-off rates and slope rates up to ~33.8 dB / GHz and ~25.7 dB / GHz are achieved, respectively. Moreover, we experimentally demonstrate highly reconfigurable filter response by simply programming the tap coefficients without changing any hardware, achieving independent tuning of all three Fano characteristic parameters including the asymmetry factor, resonance linewidth, and center frequency. These results highlight the strong potential of our approach for implementing highly reconfigurable MWP Fano filters with steep spectral transitions, which are capable of addressing diverse requirements in practical applications.

## II. Operation Principle

In this work, we investigate Fano filters exhibiting asymmetric spectral lineshapes, which were first elucidated by Fano through the well-known Fano formula as follows [20]

$$F(\varepsilon) = \frac{(q + \varepsilon)^2}{1 + \varepsilon^2}, \tag{1}$$

where $q$ is the asymmetry factor and $\varepsilon$ is the scale of reduced energy. In **Eq. (1)**, the minimum transmission reaches 0 when $\varepsilon = -q$. In spectral filter applications, ε can be further defined as

$$\varepsilon = \frac{\omega - \omega_c}{\Gamma}, \tag{2}$$

where $\omega$ is the angular frequency, $\omega_c$ is the center angular frequency, and $\Gamma$ is the resonance linewidth. By substituting $\omega = 2\pi f$, $\omega_c = 2\pi f_c$ and $\Gamma_f = \Gamma / (2\pi)$ into **Eq. (2)**, and then substituting **Eq. (2)** into **Eq (1)**, the transfer function for MW Fano filters investigated in this work can be expressed as

$$T(f) = \frac{\left(q + \frac{f - f_c}{\Gamma_f}\right)^2}{1 + \left(\frac{f - f_c}{\Gamma_f}\right)^2}, \tag{3}$$

where $f$, $f_c$, and $\Gamma_f$ are the corresponding MW frequency, center frequency, and resonance linewidth, respectively. Note that **Eq. (3)** only contains the real-valued amplitude response and does not include any phase information. Therefore, the subsequent filter designs based on **Eq. (3)** rely solely on amplitude response synthesis.

In **Fig. 1**, we plot the spectral amplitude response of Fano filters based on **Eq. (3)**. **Fig. 1(a)** illustrates the roles of three characteristic parameters, namely $q$, $\Gamma_f$, and $f_c$, in determining the spectral lineshape of the Fano filter. Here, points $A$ and $B$ mark the maximum and minimum transmission at frequencies of $f_A$ and $f_B$, respectively. Point $C$ corresponds to the center frequency of $f_c$. As can be seen, the product of $q\Gamma_f$ jointly determines the frequency offset between points $B$ and $C$, which can be expressed as

$$f_B - f_c = q\Gamma_f. \tag{4}$$

To better illustrate the relationship in **Eq. (4)**, **Figs. 1(b)**, **(c)**, and **(d)** compare the amplitude response of Fano filters with various $q$, $\Gamma_f$, and $f_c$, respectively. In each figure, all parameters other than the one being varied are kept constant. **Fig. 1(b)** shows the results for $q$ = -1 and $q$ = -4 at fixed $\Gamma_f = \Gamma_1$ and $f_c = f_1$. At $q$ = -1, the response spectrum exhibits an odd-symmetric profile, in which the transmission at the center frequency $f_c$ equals 0.5 (*i.e.*, -3 dB). According to **Eq. (4)**, it can also be obtained that $f_B - f_1 = \Gamma_1$. At $q$ = -4, the response spectrum becomes asymmetric. Point $B$ shifts to point $B$' at a higher frequency of $f_{B'}$, satisfying $f_{B'} - f_1 = 4\Gamma_1$ according to **Eq. (4)**. In addition, the transmission at $f_c$ increases and no longer equals 0.5, which is also governed by the relationship in **Eq. (4)**. **Fig. 1(c)** shows the response spectra for $\Gamma_f = \Gamma_1$ and $\Gamma_f = 2\Gamma_1$ at fixed $q$ = -1 and $f_c = f_1$. As $\Gamma_f$ increases from $\Gamma_1$ to $2\Gamma_1$, the response spectrum remains odd-symmetric, and point $B$ shifts to point $B$' at a higher frequency of $f_{B'}$. The former indicates a preserved filter symmetry, and the latter signifies an increased filtering bandwidth. **Fig. 1(d)** shows the response spectra for $f_c = f_1$ and $f_c = f_2$ at fixed $q$ = -1 and $\Gamma_f = \Gamma_1$. The filter shape and bandwidth are identical in both cases, differing only in the center frequency. This reflects the fact that variation in $f_c$ does not alter the filter shape and bandwidth, forming the basis for implementing Fano filters with tunable center frequencies.

In practical applications, another two parameters – the slope rate (SR) and the roll-off rate (ROR), are widely used to quantitatively compare the performance of Fano filters [9, 12, 21]. The SR is defined as [22]

$$\mathrm{SR} = \frac{ER}{|f_P - f_N|}, \tag{5}$$

where $ER$ is the extinction ratio, *i.e.*, the attenuation from transmission peak to notch, typically expressed in units of dB. The terms $f_P$ and $f_N$ are the frequencies corresponding to the transmission peak and notch, respectively. The ROR is defined as [4]

$$\mathrm{ROR} = \frac{AT}{|f_T - f_R|}, \tag{6}$$

where $AT$ is the attenuation from the -3 dB transmission point in the passband to a point corresponding to the reference level in the stopband. The terms $f_T$ and $f_R$ are the frequencies corresponding to these two points.

**Figs. 1(e)** and **1(f)** illustrate the definitions of SR and ROR in **Eqs. (5)** and **(6)**, respectively. For both SR and ROR, higher values indicate steeper transitions and sharper roll-off in the filter response, which is desirable for practical Fano filter applications [9, 11, 23]. The SR is more commonly used for optical resonators with Fano resonances [12-14], whereas the ROR is typically used in the context of MW filters.

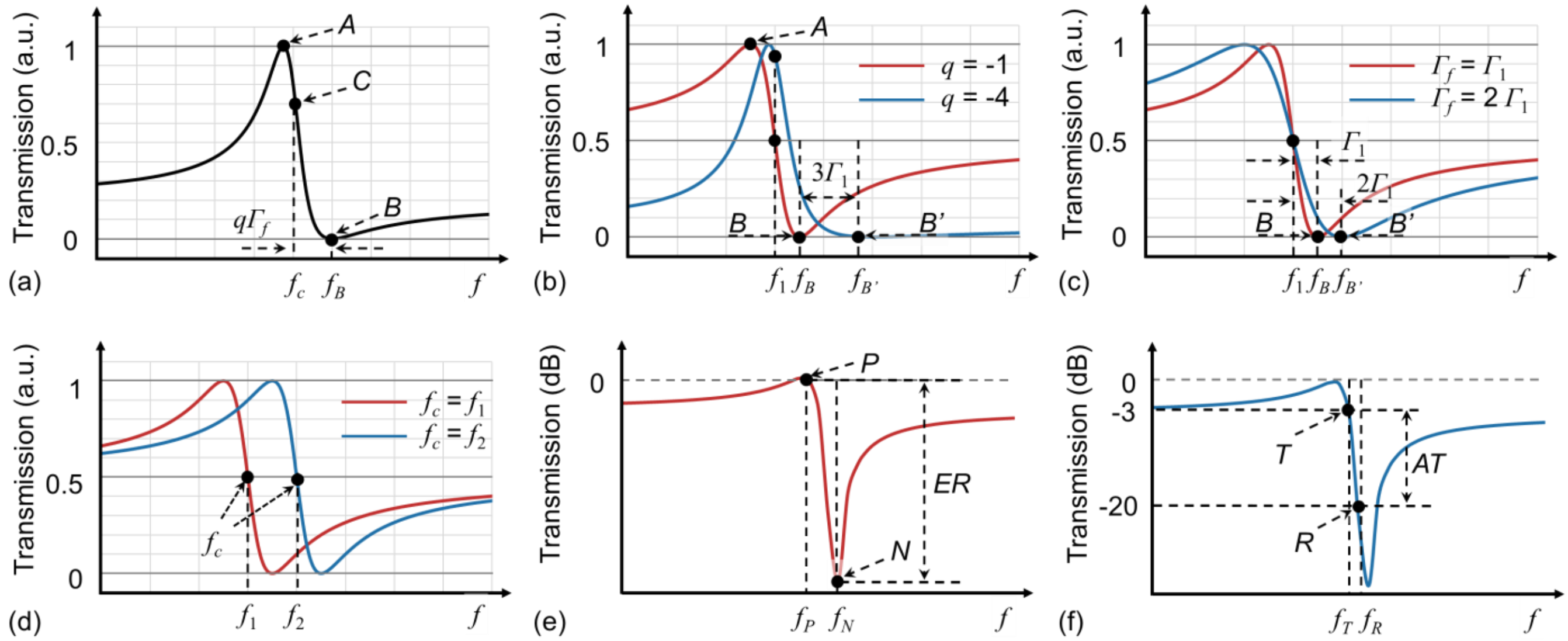

**Fig. 1.** Spectral response of Fano filters. (a) Amplitude response of a typical Fano filter. Points $A$ and $B$ mark the maximum and minimum transmission at frequencies of $f_A$ and $f_B$, respectively. Point $C$ corresponds to the center frequency of $f_c$, $q$ is the asymmetry factor, and $\Gamma_f$ is the resonance linewidth. (b) Amplitude response for $q$ = -1 and $q$ = -4 at fixed $f_c = f_1$ and $\Gamma_f = \Gamma_1$. (c) Amplitude response for $\Gamma_f = \Gamma_1$ and $\Gamma_f = 2\Gamma_1$ at fixed $q$ = -1 and $f_c = f_1$. (d) Amplitude response for $f_c = f_1$ and $f_c = f_2$ at fixed $q$ = -1 and $\Gamma_f = \Gamma_1$. (e) Illustration for the definition of slope rate (SR) for a Fano filter, where $P$ and $N$ mark the maximum and minimum transmission points at frequencies of $f_P$ and $f_N$, respectively. (f) Illustration for the definition of roll-off rate (ROR) for a Fano filter, where $T$ and $R$ mark the -3 dB transmission point in the passband and a reference point in the stopband at -20 dB transmission, respectively.

To realize the Fano filters shown in **Fig. 1**, we employ a transversal filter system widely used for MW signal processing [24-30]. **Fig. 2(a)** illustrates the operation principle of a MW transversal filter system. As the input MW signal propagates through the system, a cascade of delay elements introduces time delays between adjacent signal replicas, with each delay element providing a time delay of $\Delta t$. The delayed signal replica in each channel is weighted according to the designed tap coefficient (*i.e.*, $a_0$, $a_1$, …, $a_{M-1}$), after which the delayed and weighted signal replicas from different channels are summed to produce the final output.

**Fig. 2(b)** illustrates a transversal filter system implemented based on the MWP technology. Optical microcombs generated by a compact integrated microring resonator (MRR) is employed as a multiwavelength source for the transversal filter system. Replicas of the input MW signal are generated by modulating it onto optical carriers at different wavelength channels. The delay between adjacent wavelength channels is provided by the dispersion of a spool of single-mode fiber (SMF), and the tap coefficient in each channel is assigned using a programmable optical spectral shaper (OSS). Finally, the delayed and weighted signal replicas are summed upon photodetection to generate a MW signal as the system output.

The output MW signal $s(t)$ from the MWP transversal filter system in **Fig. 2(b)** can be expressed as

$$s(t) = f(t) * h(t) = \sum_{n=0}^{M-1} a_n f(t - n\Delta t), \tag{7}$$

where $M$ is the total number of tap coefficient, $a_n$ ($n$ = 0, 1, 2, …, $M$ - 1) is the tap coefficient of the $n_{th}$ tap, $\Delta t$ is the delay time between adjacent taps and $f(t)$ denoting the input MW signal. The temporal impulse response $h(t)$ in **Eq. (7)** can be given by

$$h(t) = \sum_{n=0}^{M-1} a_n \delta(t - n\Delta t), \tag{8}$$

where $\delta(t)$ is the unit impulse function. By applying the Fourier transform to **Eq. (8)**, the spectral transfer function of the MWP transversal filter system can be written as

$$H(\omega) = \sum_{n=0}^{M-1} a_n e^{-j\omega n\Delta t}, \tag{9}$$

**Eq. (9)** represents a typical transfer function for the transversal filter systems [31], which exhibits finite impulse response (FIR), and diverse filtering functions can be realized by designing the corresponding tap coefficients $a_n$ ($n$ = 0, 1, 2, …, $M$ - 1). This provides the basis for implementing Fano filters shown in **Fig. 1**, enabling reconfigurable Fano filters with different asymmetry factor $q$, resonance linewidth $\Gamma_f$, or center frequencies $f_c$ to be realized simply through programing the tap coefficients $a_n$ ($n$ = 0, 1, 2, …, $M$ - 1).

It is also worth noting that due to the FIR nature of transversal filter systems, deviations arise between the ideal filter response and that realized based on the transfer function in **Eq. (9)** for target filters with infinite impulse response (IIR), including Fano filters investigated in this work. These deviations can also degrade the SR and ROR of Fano filters realized based on the MWP transversal filter system. However, these deviations decrease as the tap number $M$ increases and becomes negligible when $M$ is sufficiently large.

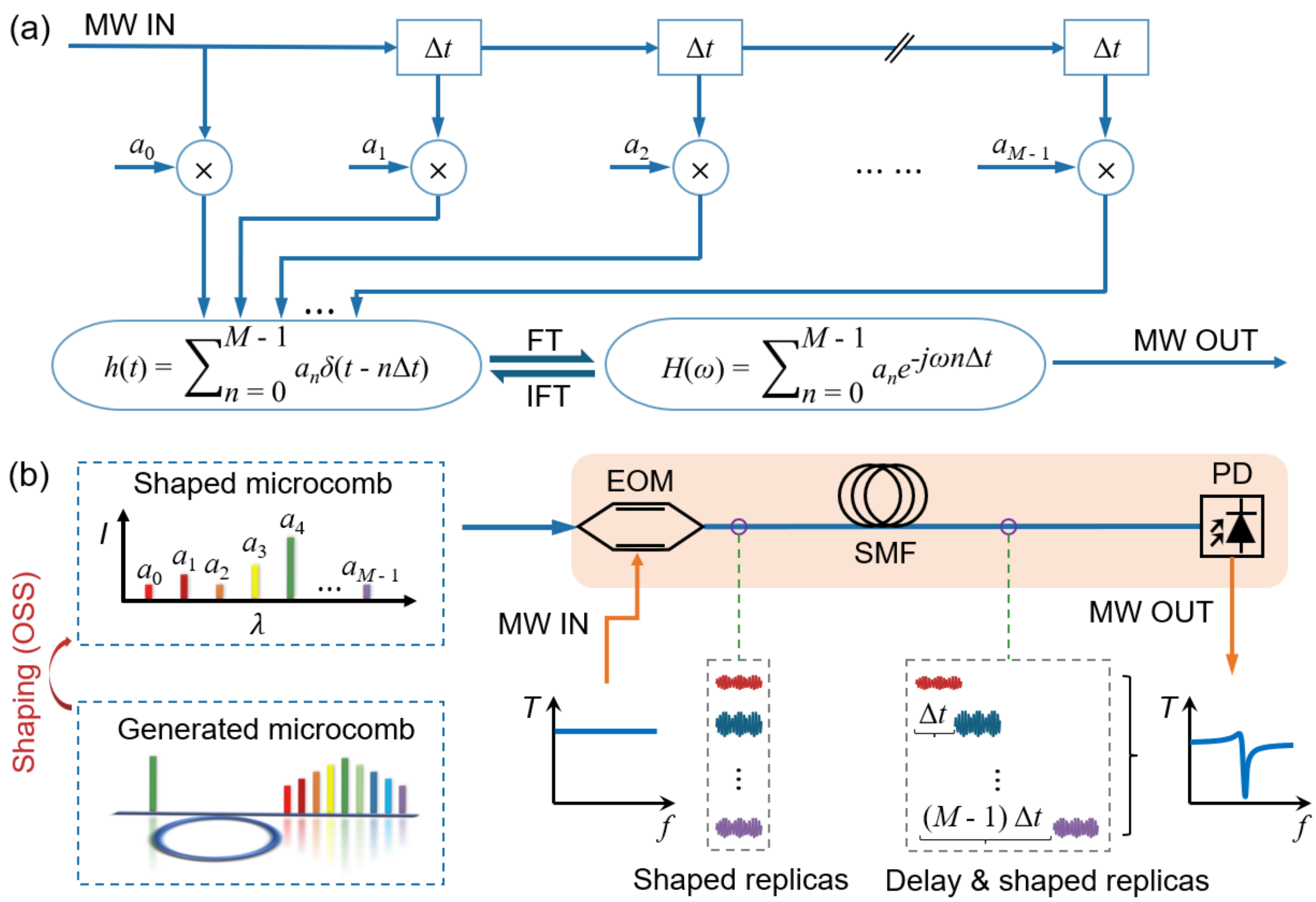

**Fig. 2.** Transversal filter system. (a) Schematic illustration for the operation principle of a microwave (MW) transversal filter system, where $\Delta t$ is the time delay between adjacent channels, and $a_0$, $a_1$, …, $a_{M-1}$ are the tap coefficients in each channel. FT: Fourier transform. IFT: inverse Fourier transform. (b) Schematic diagram and processing flow of a microwave photonic (MWP) transversal filter system with an optical microcomb source. OSS: Optical spectral shaper. EOM: Electro-optic modulator. PD: photodetector.

Therefore, a large number of wavelength channels are required in practical MWP transversal filter systems to minimize these deviations and increase the SR and ROR performance. For brevity, the filters realized based on **Eq. (9)** are referred to as MWP Fano filters throughout this paper, although strictly speaking, they are synthesized filters exhibiting a Fano-like amplitude spectral response in the MW domain, rather than those based on directly mapping the Fano resonances generated by IIR optical filters [9-11]. Unlike the IIR optical filters that generate Fano responses through complex phase interference, our FIR transversal filter can directly synthesize the desired Fano-shaped amplitude response. In addition, using amplitude information alone does not affect the applicability of our filters to frequency discrimination, where the MW frequency can be identified through amplitude-to-frequency mapping using the steep Fano amplitude slopes, similar to those in Refs. [32, 33]. The use of a real-valued amplitude response could also reduce sensitivity to phase noise and fluctuations caused by vibrations or thermal drifts, thereby improving system stability and supporting the maintenance of high RORs and SRs.

In **Fig. 2(b)**, employing optical microcombs as the multiwavelength source offers significant advantages over conventional discrete laser arrays [34-36] and fiber Bragg grating arrays [37-39], which are constrained in the number of available taps because the system size, power consumption, and complexity increase dramatically with the tap number. In addition, compared with laser frequency combs generated by electro-optic modulation or mode-locked fiber lasers [40-42], optical microcombs provide large comb spacings owing to the small volume of the micro-resonators, enabling wide Nyquist bands between wavelength channels and hence broad operation bandwidths for the MWP transversal filters.

In **Eq. (9)**, the time delay $\Delta t$ provided by the SMF in **Fig. 2(b)** can be given by [43]

$$\Delta t = L \cdot D \cdot \Delta\lambda\,, \tag{10}$$

where $\Delta\lambda$ denotes the comb spacing, $D$ is the dispersion parameter of the SMF, and $L$ is the length of the SMF. Since the MWP transversal filter has finite impulse response, it exhibits a periodic spectral response with a free spectral range (FSR) of $FSR_{MW}$ = 1 / $\Delta t$. This sets an upper limit for the operation bandwidth (OBW) of the microcomb-based MWP transversal filter system, which is equal to $FSR_{MW}$ / 2. In addition, according to the Nyquist sampling theorem, a bandwidth-limited continuous-time signal must be sampled at a rate exceeding twice its highest frequency component to avoid aliasing. This constraint therefore sets another upper limit on the operation bandwidth of the microcomb-based MWP transversal filter system, which is equal to half of the comb spacing (*i.e.*, $\Delta\lambda$ / 2). Considering these above, the operation bandwidth (OBW) of a microcomb-based MWP transversal filter can be expressed as:

$$\text{OBW} = \min\{\Delta\lambda / 2\,,\ FSR_{MW} / 2\}. \tag{11}$$

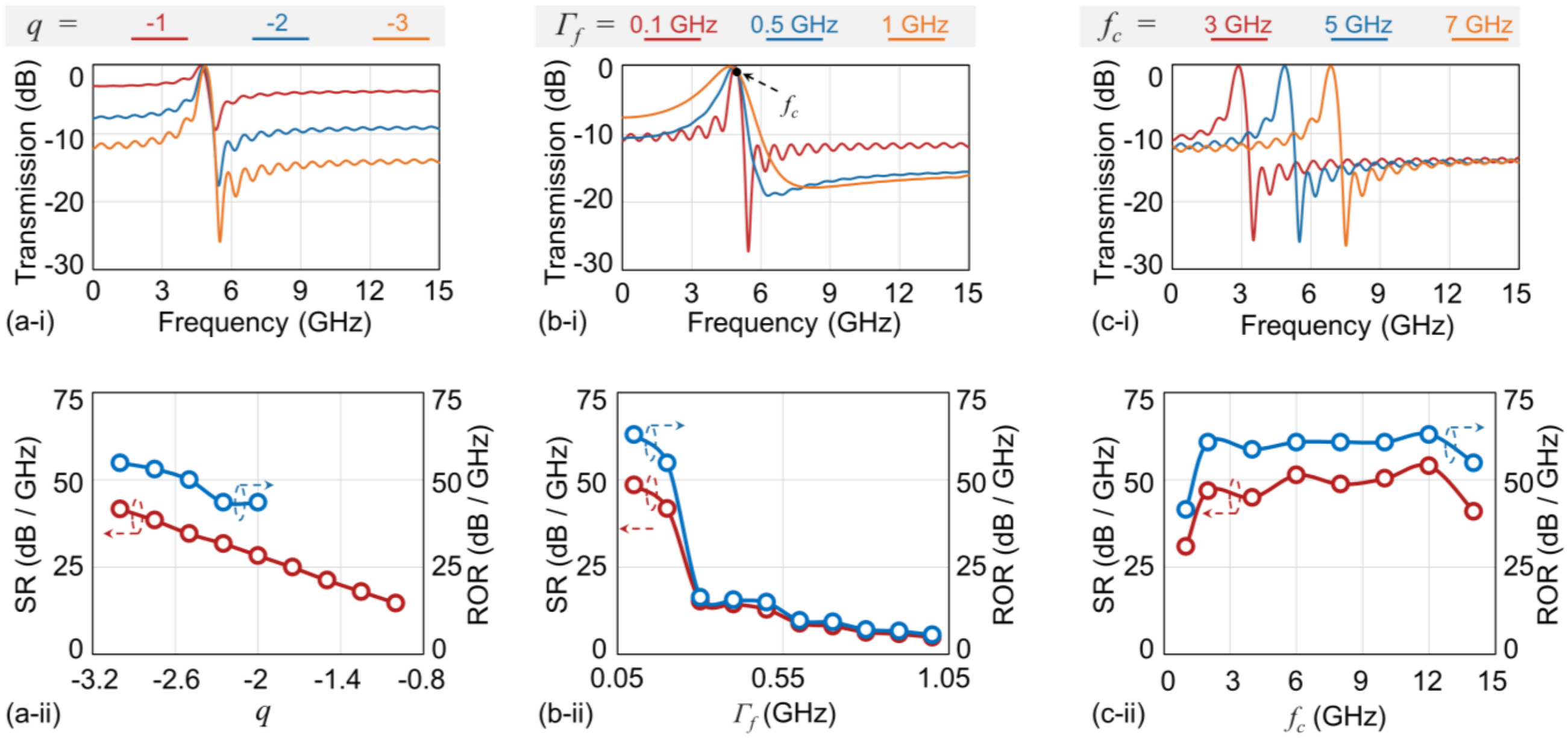

**Fig. 4.** Simulated response of Fano filters with various asymmetry factors $q$, resonance linewidth $\Gamma_f$, and center frequencies $f_c$, realized using a microcomb-based MWP transversal filter system. (a-i) Simulated spectral response for $q$ = -1, -2, and -3 at fixed $f_c$ = 5 GHz and $\Gamma_f$ = 0.2 GHz. (a-ii) Calculated SR and ROR versus $q$. (b-i) Simulated spectral response for $\Gamma_f$ = 0.1, 0.5, and 1 GHz at fixed $f_c$ = 5 GHz and $q$ = -3. (b-ii) Calculated SR and ROR versus $\Gamma_f$. (c-i) Simulated spectral response for $f_c$ = 3, 5, and 7 GHz at fixed $q$ = -3 and $\Gamma_f$ = 0.2 GHz. (c-ii) Calculated SR and ROR versus $f_c$.

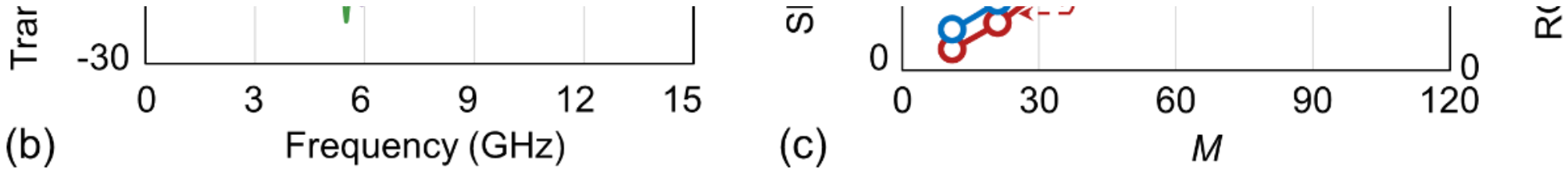

**Fig. 3.** Simulated spectral response of Fano filters with $q$ = -3, $\Gamma_f$ = 0.2 GHz, and $f_c$ = 5 GHz, realized using a microcomb-based MWP transversal filter system. (a-i) – (a-iv) Designed tap coefficients for Fano filters with various tap numbers of $M$ = 11, 21, 41, and 81, respectively. (b) Simulated amplitude response for the Fano filters in (a). (c) Slope rate (SR) and roll-off rate (ROR) versus $M$, which are calculated based on the results in (b).

**Fig. 3** shows simulated response of Fano filters realized using the microcomb-based MWP transversal filter system in **Fig. 2(b)**. In our simulation, the parameters in **Eq. (3)** were set to $q$ = -3, $\Gamma_f$ = 0.2 GHz, and $f_c$ = 5 GHz. **Fig. 3(a)** shows the designed tap coefficients across different wavelength channels. Here we show the results for different tap numbers of $M$ = 11, 21, 41 and 81. The tap coefficients (*i.e.*, $a_0$, $a_1$, $a_2$, …, $a_{M-1}$), including both positive and negative values, were calculated by applying inverse Fourier transform (IFT) to the target transfer function in **Eq. (3)**. As can be seen, taps with identical coefficients are symmetrically distributed about the center tap. As a result, each pair of taps with equal amplitudes and opposite phase terms combines to produce a constant time delay, leading to a linear phase response with a constant group delay. In addition, the tap coefficients near the center exhibit larger absolute values and dominate the synthesis of the filter response, whereas those on both sides have smaller absolute values and contribute less significantly. This feature is determined by the temporal impulse response of ideal Fano filters.

**Fig. 3(b)** shows simulated amplitude response of Fano filters corresponding to the tap distributions in **Fig. 3(a)**. For the ideal Fano response shown in **Fig. 1(a)**, the minimum transmission is zero, resulting in an infinite ER (in dB). In the finite-tap transversal filter implementation, however, the ER becomes finite and increases with the tap number $M$, due to impulse-response truncation and finite-tap approximation of the ideal response. For $M$ = 11, the main asymmetric Fano profile is reproduced. However, the notch remains relatively shallow, and the spectrum outside the resonance exhibits noticeable ripples, leading to limited SR and ROR. As $M$ increases, the response exhibits a more distinct peak to notch feature, together with suppressed ripples as well as improved SR and ROR. For $M$ = 81, a well-defined Fano lineshape with only minor ripples is observed, achieving a high *ER* of ~25.9 dB.

To quantitatively analyze the influence of tap number on the Fano filters, **Fig. 3(c)** plots the SR and ROR of the Fano filters versus tap number $M$. As expected, both SR and ROR increase monotonically with increasing $M$, confirming that a larger tap number yields a steeper roll-off for the Fano filters. The simulation results show that increasing $M$ from 11 to 81 improves the SR from ~1.94 to ~41.8 dB / GHz and the ROR from ~6.56 to ~54.8 dB / GHz. As $M$ increases beyond 81, further improvements in the SR and ROR become marginal. Considering this, a maximum tap number of $M$ = 81 is chosen for our following experimental demonstrations.

**Fig. 4** shows simulated response of reconfigurable Fano filters with various asymmetry factors $q$, resonance linewidth $\Gamma_f$ and, center frequencies $f_c$, realized using the microcomb-based MWP transversal filter system in **Fig. 2(b)**. **Fig. 4(a-i)** shows the results for $q$ = -1, -2 and -3 at fixed $f_c$ = 5 GHz and $\Gamma_f$ = 0.2 GHz. As $q$ changes from -1 to -3, the filter shape becomes increasingly asymmetric, together with an increase in the *ER*. **Fig. 4(a-ii)** shows the calculated SR and ROR versus $q$ over a range of $-1 \leq q \leq -3$. The stopband reference level was set to -20 dB for the ROR calculation according to **Eq. (6)**, and we did not plot the ROR values for $q > -2$ because the stopband response remains above this reference level. Both the SR and the ROR increase as the absolute value of $q$ increases – consistent with the trend in **Fig. 1(b)**. **Fig. 4(b-i)** shows the results for $\Gamma_f = 0.1$, 0.5, and 1 GHz at fixed $f_c = 5$ GHz and $q = -3$. According to the relationship governed by **Eq. (4)**, the frequency offset from $f_c$ to the minimum transmission point equals $3\Gamma_f$. Therefore, the filter exhibits the steepest roll-off at the lowest $\Gamma_f = 0.1$ GHz. The corresponding SR and ROR in **Fig. 4(b-ii)** decrease with increasing $\Gamma_f$, further confirming this trend. **Fig. 4(c-i)** shows the results for $f_c = 3$, 5, and 7 GHz at fixed $q = -3$ and $\Gamma_f = 0.2$. The filter shape and bandwidth remain unchanged, with only the center frequency position shifting – consistent with the results in **Fig. 1(d)**. This is also reflected in the nearly flat SR and ROR curves in **Fig. 4(c-ii)**. We also note that the SR and ROR curves exhibit slight degradation on both sides of the operating band. This is mainly attributed to the finite-tap approximation of the discrete-time transversal filter. At low $f_c$, the filter response approaches the resolution limit. Whereas at high $f_c$, the frequency components approach the Nyquist boundary, making the response increasingly susceptible to periodic boundary effects and truncation-induced ripples. These effects elevate the notch floor and distort the Fano filter shape, leading to reduced SR and ROR.

## III. EXPERIMENTAL SETUP

Based on the theory in **Section II**, we performed experimental demonstrations for reconfigurable MWP Fano filters based on optical microcombs. **Fig. 5** shows a schematic of the experimental setup. A tunable continuous-wave (CW) laser was amplified by an erbium-doped fiber amplifier (EDFA) and served as the pump light. The polarization of the pump light was adjusted by a polarization controller (PC) before being coupled into a nonlinear MRR to generate optical microcombs. A temperature controller (TC) was employed to stabilize the chip temperature, thereby ensuring long-term stable microcomb generation. The initially generated microcomb exhibited non-uniform comb line powers, with higher powers in primary comb lines and lower powers in other wavelength channels. Therefore, an OSS was employed to flatten the comb lines.

The flattened optical microcomb was first amplified by another EDFA and then sent to the transversal filter module consisting of a PC, an EOM, a spool of single-mode fibre (SMF), another OSS, and a balanced photodetector (BPD). The EOM, SMF, and OSS served the same functions as those discussed in **Fig. 2(b)**. The PC was employed to optimize the modulation efficiency of the polarization-sensitive EOM. Compared with the PD in **Fig. 2(b)**, the BPD performed a similar function but was connected to the two complementary output ports of the OSS. This separated the wavelength channels into two groups and introduced a π phase shift between them, thus allowing both positive and negative tap coefficients. To minimize the time-delay mismatch between the two tap groups, two fibers of nearly identical lengths were employed. A vector network analyzer (VNA, Rohde &

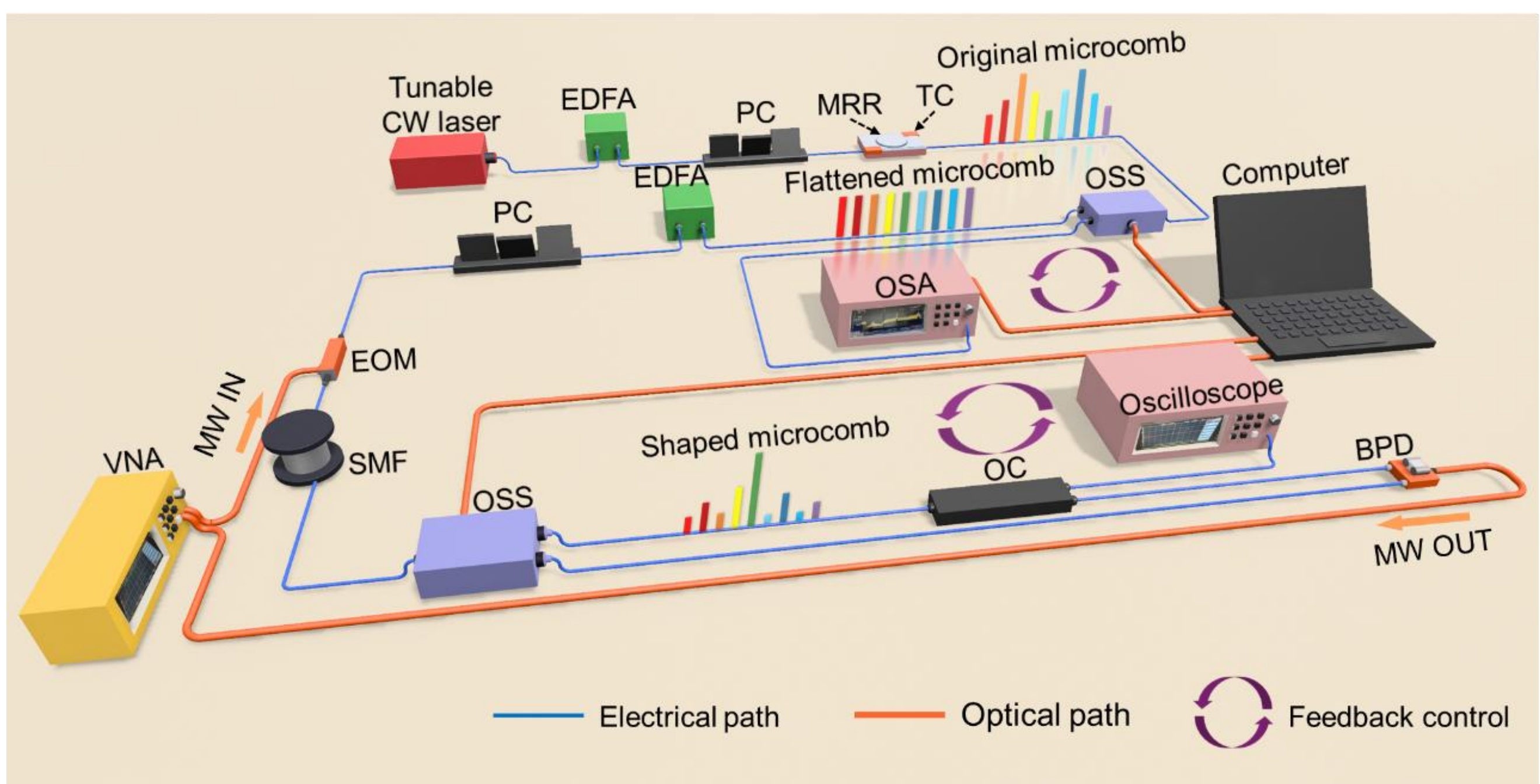

**Fig. 5.** Schematic of the experimental setup for demonstrating reconfigurable microcomb-based MWP Fano filters. CW laser: continuous-wave laser. EDFA: erbium-doped fiber amplifier. PC: polarization controller. MRR: micro-ring resonator. TC: temperature controller, OSS: optical spectral shaper. OSA: optical spectrum analyzer. EOM: electro-optic modulator. SMF: single-mode fiber. OC: optical coupler. BPD: balanced photodetector. VNA: vector network analyzer.

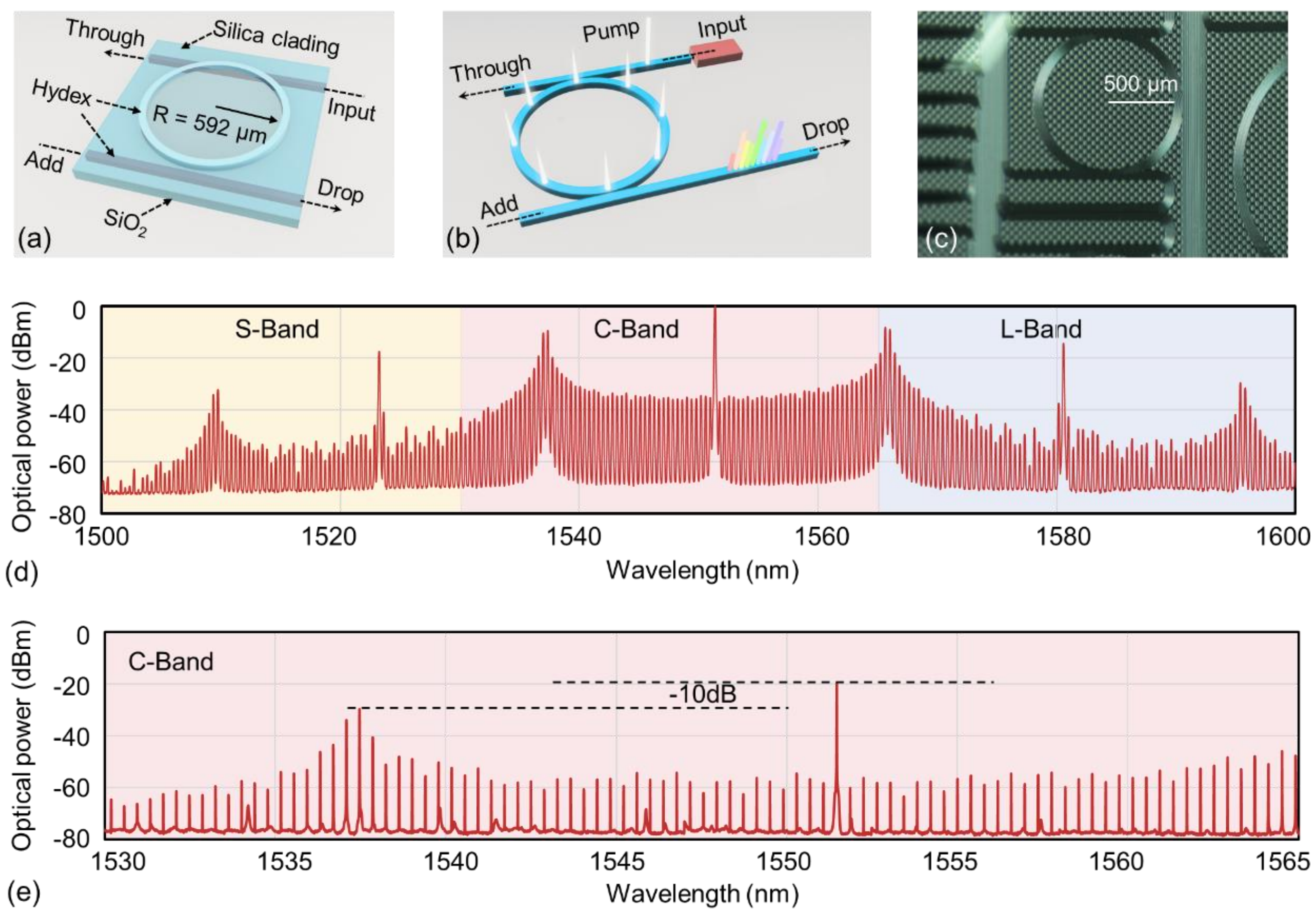

**Fig. 6.** Optical microcomb generation. (a) Schematic of a doped silica MRR used for optical microcomb generation. (b) Schematic illustration of optical microcomb generation from the MRR in (a). (c) Microscope image of the fabricated doped silica MRR. (d) Optical spectrum of soliton crystal microcomb generated by the MRR in (c). (e) Zoom-in view of the spectrum in (d) within telecom C band.

Schwarz) was employed to measure the response of the MWP transversal filter system. An optical spectrum analyzer (OSA, Anritsu) was used to measure the power of the shaped comb lines. We also employed a two-stage feedback control strategy, including synergic spectral power reshaping and impulse response reshaping as detailed in Ref. [44], to minimize the errors induced by imperfect response of experimental components and improve the accuracy of comb line shaping.

In our experimental demonstration, soliton crystal microcombs generated by an integrated doped silica MRR were employed as the microcomb source. **Fig. 6(a)** shows a schematic of the doped silica MRR. The radius of the MRR is ~592 μm. Two bus waveguides are coupled to the central micro-ring, forming a four-port device. **Fig. 6(b)** illustrates the generation of soliton crystal optical microcombs from the MRR in **Fig 6(a)**. The drop port of the MRR is employed as the output port due to its inherent filtering effect that suppresses noise in the generated optical microcombs. Soliton crystal microcombs are a distinct class of optical microcomb in which multiple co-circulating solitons spontaneously self-organize into an ordered, crystal-like pattern along the MRR [29, 30].

**Fig. 6(c)** shows a microscope image of the fabricated MRR. The MRR was fabricated on a doped silica platform using complementary metal-oxide-semiconductor (CMOS) compatible processes [45, 46]. Doped silica films with a refractive index of ~1.7 at 1550 nm were first deposited by plasma-enhanced chemical vapor deposition (PECVD), patterned by deep ultraviolet (UV) photolithography, and etched via reactive ion etching (RIE) to form low roughness waveguides. A silica upper cladding layer with a refractive index of ~1.44 at 1550 nm was then deposited. The doped silica platform offers low linear propagation loss of ~0.06 dB · $cm^{-1}$, a moderately high optical nonlinear parameter of ~233 $W^{-1}$ · $km^{-1}$, and negligible nonlinear optical loss even at intensities up to 25 GW · $cm^{-2}$. The fabricated MRR had a high quality (Q) factor of ~1.9 million and an FSR of ~0.4 nm (*i.e.*, ~49 GHz). After packaging the MRR with fiber pigtails at the input and output ports, the coupling loss was less than 1 dB per facet.

**Fig. 6(d)** shows the optical spectrum of soliton crystal microcomb generated by the MRR in **Fig. 6(c)**. The comb spacing was ~0.4 nm (*i.e.*, ~49 GHz). The optical microcomb was generated from the MRR by amplifying a CW pump to ~32.1 dBm and sweeping it from shorter to longer wavelengths across a TE-polarized resonance near 1551.3 nm. As the detuning between the pump and the cold-cavity resonance decreased, the intracavity power increased and exceeded the threshold for modulation instability (MI), which initiated MI oscillations [47]. Primary comb lines were then generated, with the initial comb spacing set by the MI gain peak that is mainly governed by the cavity dispersion and the intracavity power. By further increasing the pump detuning, featured fingerprint-like spectra for soliton crystal microcombs were observed, showing agreement with those reported in Refs. [48-50]. Compared with dissipative Kerr solitons [4, 43], soliton crystal microcombs experience minimal intracavity energy variation during their formation, which enables simple and robust initiation by adiabatically sweeping the pump wavelength via manual detuning [4, 30]. **Fig. 6(e)** shows a zoom in view of the microcomb spectrum in **Fig. 6(d)** within the telecom C band (*i.e.*, 1530 nm – 1565 nm). Owing to the relatively small FSR of the MRR, a total of 90 comb lines fall within the C band, providing sufficient taps for achieving high accuracy of the transversal filter system.

For experimental setup in **Fig. 5**, the length and dispersion of the SMF were $L$ = 4.8 km and $D$ = 17.4 ps · $nm^{-1}$ · $km^{-1}$, respectively. These, together with the comb spacing of $\Delta\lambda$ = 0.4 nm in **Fig. 6(d)**, result in a time delay of $\Delta t = L \cdot D \cdot \Delta\lambda$ = ~0.033 ns between adjacent wavelength channels and an MW FSR of $FSR_{MW} = 1 / \Delta t$ = ~30 GHz. According to **Eq. (8)**, the OBW of the practical microcomb-based MWP transversal filter system is OBW = min $\{\Delta\lambda / 2, FSR_{MW} / 2\}$ = min {~24.5 GHz, ~15 GHz} = ~15 GHz.

## IV. Experimental results

By using the experimental setup discussed in **Section III**, we performed demonstrations for reconfigurable MWP Fano filters. In this section, we present and discuss the experimental results, including the filter response for various tape numbers and reconfigurable filter response with varying characteristic parameters.

**Fig. 7** shows the results for MWP Fano filters with various tap numbers of $M$ = 11, 21, 41, and 81. For comparison, the characteristic parameters of the Fano filters were fixed at $f_c$ = 5 GHz, $q$ = -3, and $\Gamma_f$ = 0.2 GHz. **Figs. 7(a-i) – (a-iv)** show the designed tap coefficients (circles) and measured optical spectra of shaped comb lines (solid lines) for $M$ = 11, 21, 41, and 81, respectively. The yellow circles and lines represent positive tap coefficients, whereas the orange circles and lines correspond to negative tap coefficients. As the tap number increases, more comb lines of the generated microcomb were utilized for the transversal filter, enabling a closer approximation to the ideal Fano filter transfer function in **Eq. (3)**.

**Fig. 7(b)** shows measured amplitude response of the Fano filters corresponding to the results in **Fig. 7(a)**, which was measured by the VNA in **Fig. 5**. At $M$ = 11, the filter response exhibits a low $ER$ of ~7.19 dB, together with a gradual spectral roll-off. As $M$ increases, the $ER$ increases and the roll-off becomes steeper, in agreement with the simulation results in **Fig. 3(b)**. **Fig. 7(c)** shows the calculated SR and ROR versus $M$ based on the results in **Fig. 7(b)**. To facilitate comparison of the experimental results, the stopband reference level was set to -13 dB when calculating the ROR based on **Eq. (6)**, and the ROR was not evaluated in cases where the minimum transmission exceeded this reference level. As $M$ increases from 11 to 81, the $SR$ increases monotonically from ~1.4 to ~25.7 dB/GHz. For $M$ > 41, the increase in SR becomes marginal, and a similar trend was observed for the ROR reaching a maximum value of ~33.8 dB / GHz at $M$ = 81. These are consistent with the simulation results in **Fig. 3(c)**, further confirming that the taps near the center contribute more significantly to shape the filter response. Considering this, we choose a fixed tap number of $M$ = 41 for the demonstration of reconfigurable MWP Fano filters in **Fig. 8**.

**Fig. 8** shows the results for MWP Fano filters with reconfigurable characteristic parameters. In our experiments, we demonstrated independent tuning of the asymmetry factor $q$, resonance linewidth $\Gamma_f$, and center frequency $f_c$, which was achieved by programing the tap coefficients allocated to different wavelength channels without changing any hardware. For comparison, only one characteristic parameter varied and the other two remained unchanged.

**Fig. 8(a-i)** shows the measured amplitude response of the Fano filters for $q$ = -1, -2, and -3 at $f_c$ = ~5 GHz and $\Gamma_f$ = ~0.2 GHz. As $q$ changed from -1 to -3, the response became more asymmetric, accompanied by an increase in the $ER$. These trends are consistent with those observed in **Fig. 4(a-i)**. **Fig. 8(a-ii)** shows the calculated SR and ROR versus $q$ based on the results in **Fig. 8(a-i)**. We do not plot the ROR values for $q$ < -1 because the stopbands remain above the -13 dB reference level. High SR and ROR values are achieved at $q$ = -3, and the $SR$ decreases from ~22.4 to ~12.3 dB / GHz as $q$ varies from -3 to -1. The latter exhibits a trend that agrees well with the simulation results in **Fig. 4(a–ii)**.

**Fig. 8(b-i)** shows the measured amplitude response for $\Gamma_f$ = ~0.2, ~0.8, and ~2.5 GHz at fixed $f_c$ = ~5 GHz and $q$ = -3. As $\Gamma_f$ increased, the spectral interval between the resonance peak and notch widened, resulting in a degraded spectral roll-off between them. These trends are consistent with those observed in **Fig. 4(b-i)**. **Fig. 8(b-ii)** shows the calculated $SR$ and $ROR$ versus $\Gamma_f$ based on the results in **Fig. 8(b-i)**. Both the $SR$ and $ROR$ decrease with increasing $\Gamma_f$ – in agreement with

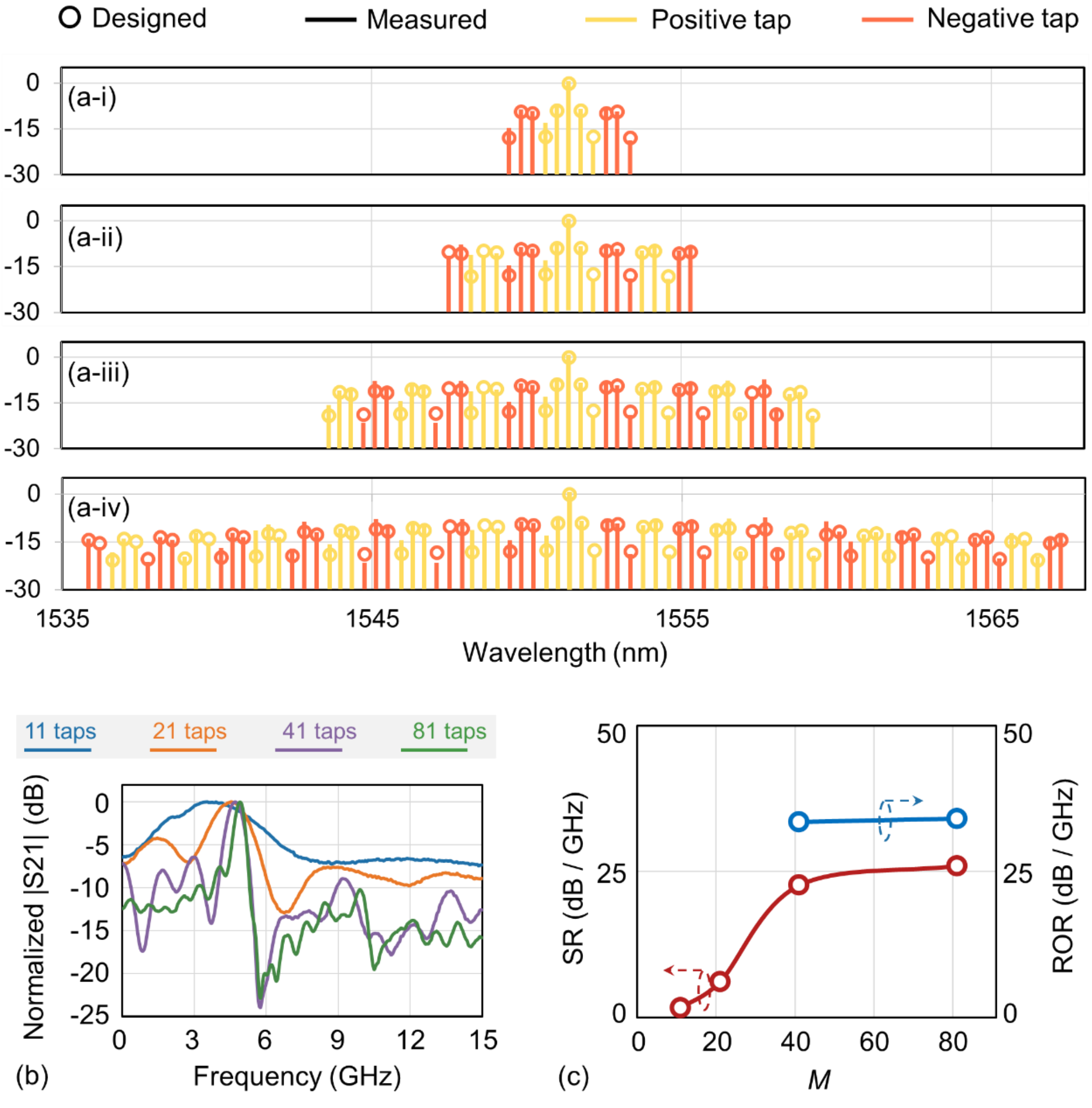

**Fig. 7.** Experimental results of MWP Fano filters with various tap numbers of $M$ = 11, 21, 41, and 81. (a) Designed tap coefficients (circles) and measured optical spectra of shaped comb lines (solid lines) for (i) $M$ = 11, (ii) $M$ = 21, (iii) $M$ = 41, and (iv) $M$ = 81, where yellow circles and lines represent positive tap coefficients, and orange circles and lines correspond to negative tap coefficients. (b) Measured MW amplitude response corresponding to the results in (a). (c) SR and ROR versus $M$ calculated based on the results in (b). In (a) – (c), the Fano characteristic parameters are $q$ = -3, $\Gamma_f$ = 0.2 GHz, and $f_c$ = 5 GHz.

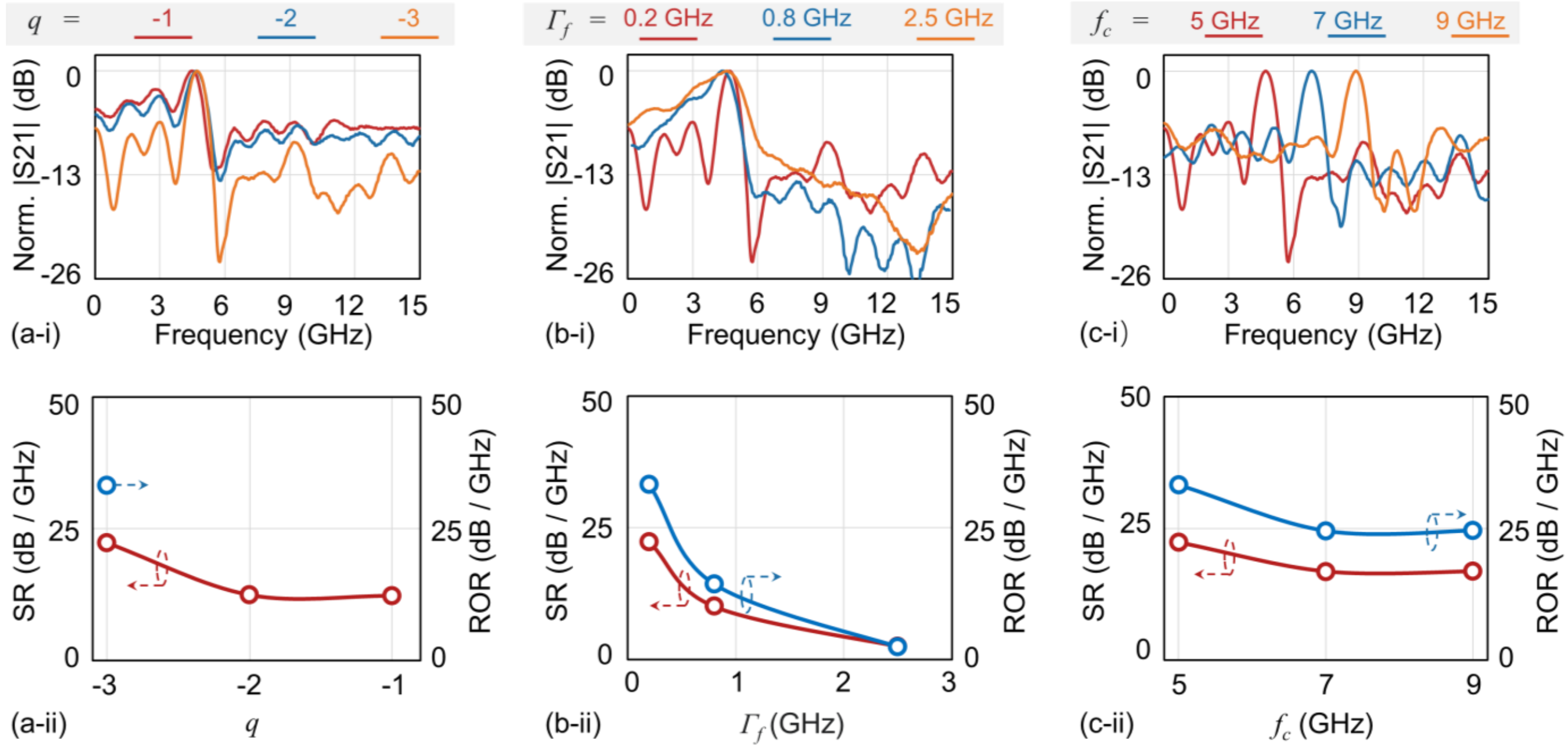

**Fig. 8.** Experimental results of reconfigurable MWP Fano filters. (a-i) Measured MW amplitude response for $q$ = -1, -2 and -3 at fixed $f_c$ = 5 GHz and $\Gamma_f$ = 0.2 GHz. (a-ii) Calculated SR and ROR based on the results in (a-i). (b-i) Measured MW amplitude response for $\Gamma_f$ = 0.2, 0.8 and 2.5 GHz at fixed $f_c$ = 5 GHz and $q$ = -3. (b-ii) Calculated SR and ROR based on the results in (b-i). (c-i) Measured MW amplitude response for $f_c$ = 5, 7 and 9 GHz at fixed $q$ = -3 and $\Gamma_f$ = 0.2 GHz. (c-ii) Calculated SR and ROR based on the results in (c-i).

the trends for the simulation results in **Fig. 4(b-ii)**.

**Fig. 8(c-i)** shows the measured response for $f_c$ = ~3, ~5, and ~7 GHz at fixed $q$ = -3 and $\Gamma_f$ = ~0.2 GHz. As $f_c$ increased, the filter shape and bandwidth remain largely unchanged, with only the center frequency shifting towards higher frequencies – consistent with the simulation results in **Fig. 4(c-i)**. **Fig. 8(c-ii)** shows the calculated *SR* and *ROR* versus $\Gamma_f$ based on the results in **Fig. 8(c-i)**. Both *SR* and *ROR* exhibit minimal changes as $f_c$ increased, agreeing with the simulation results in **Fig. 4(c-ii)**. The minor fluctuations are mainly attributed to the non-ideal response of the experimental instruments, as will be elaborated subsequently

## V. Discussion

This work represents the first demonstration of using a microcomb-based transversal filter system to realize MWP Fano filters. Compared with previous works on MWP Fano filters realized through direct mapping of optical resonator responses [9, 21], our work opens up a new technical route that offers several unique advantages.

First, by leveraging a large number of wavelength channels for response synthesis, our system enables unprecedented reconfigurability through independent tuning of all the Fano characteristic parameters, including the asymmetry factor, resonance linewidth, and center frequency. This significantly improves the filter reconfigurability, application flexibility, and operational range, and such a capability has not been demonstrated in any previous report. In contrast, previous approaches based on direct mapping of optical resonator responses at a single wavelength typically offer limited flexibility, as these characteristic parameters are often governed by the same set of device structural parameters and therefore cannot be independently controlled.

Second, our MWP Fano filters can readily synthesize high-order filter responses without relying on precise wavelength alignment. This enables excellent performance metrics, achieving RORs and SRs of up to ~33.8 dB / GHz and ~25.7 dB / GHz, respectively, which are substantially higher than the reported ROR values of ~ 7 dB / GHz [51] and ~10.2 dB / GHz [52], and SR values of ~3 dB / GHz [9] and 8 dB / GHz [21] in previous studies. These improvements are critical for enhancing the sensitivity of frequency discrimination based on MWP Fano filters.

Finally, our system offers high operational stability. For MWP Fano filters based on direct mapping of optical resonator responses, long-term operation is constrained by the need for precise wavelength alignment, which is susceptible to resonance wavelength shifts caused by thermal effects, particularly for Fano resonances with steep spectral slopes [11, 15, 23]. In contrast, our transversal filter system does not rely on wavelength alignment and therefore offers improved operational stability. Moreover, the soliton crystal microcomb employed in this work has demonstrated stable operation for ~66 hours in optical communication experiments [53]. Except for the microcomb source, all other components are commercially available and with demonstrated high stability. In **Fig. 9**, we show the S21 response measured over extended periods of 1- 3 hours. As observed, the response remains nearly unchanged in shape, demonstrating the high stability of our system.

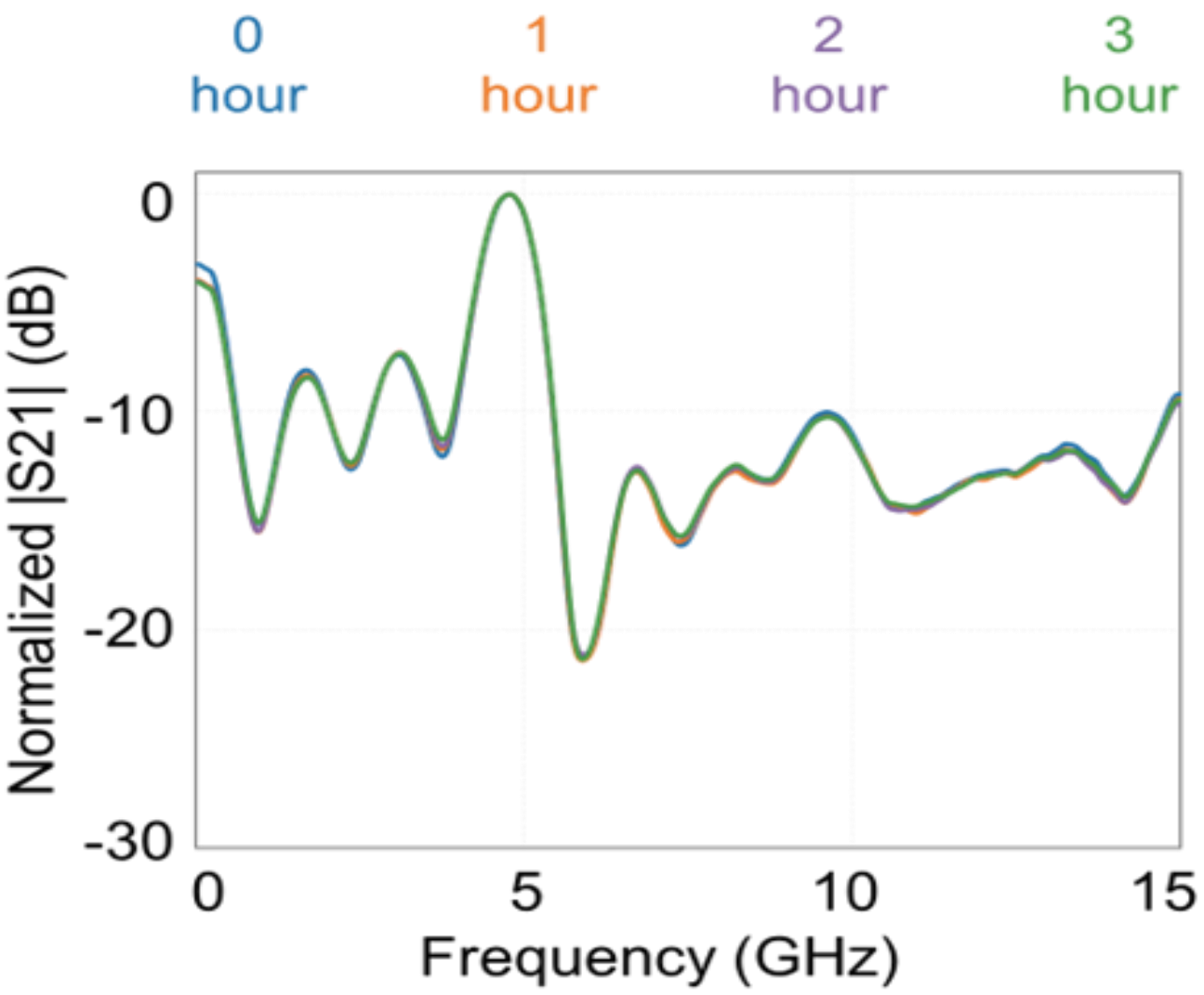

Fig. 9. Measured S21 response over extended periods of 1- 3 hours. The Fano characteristic parameters are $q$ = -3, $\Gamma_f$ = 0.2 GHz, and $f_c$ = 5 GHz.

The discrepancies between the filter response obtained in our experiments and the ideal Fano filter response in **Fig. 1** arise from two sources, namely, theoretical approximation and imperfect response of a practical system. The former refers to the theoretical approximation of a filter with an infinite impulse response (which corresponds to infinite tap number) using a practical transversal filter system with a finite tap number. In our case, the discrepancies induced by theoretical approximation were largely mitigated by employing a sufficiently large tap number $M$ (up to 81, as discussed in **Fig. 3)** and ensuring that the filter operates within the OBW (~15 GHz, as defined in **Eq. (11)**). Therefore, the discrepancies were mainly attributed to experimental imperfections, such as the noise of microcomb, chirp of the EOM, high-order dispersion of the SMF, shaping errors of the OSS, noise of the BPD, and time-delay mismatch between the positive and negative tap groups.

**Fig. 10(a)** compares the simulated Fano filter response in **Fig. 3(b)** with the experimentally measured response in **Fig. 7(b)**, and their differences are induced by hardware imperfections. To provide a quantitative comparison, **Fig. 10(b)** shows the ERs, RORs, and SRs extracted from **Fig. 10(a)**. As discussed previously, theoretical finite-tap approximation introduces discrepancies that reduce the infinite ER (in dB) of the ideal Fano filter response to finite ERs of the simulated response in **Fig. 10(a)**. Compared with the simulated response, the experimentally measured response exhibits even shallower notches and larger fluctuations. For example, at $M = 81$, the simulated ER, ROR, and SR are ~25.9 dB, ~54.8 dB/GHz, and ~41.7 dB/GHz, respectively, whereas the corresponding measured values decrease to ~23.8 dB, ~33.8 dB/GHz, and ~25.7 dB/GHz.

We also performed simulations to investigate the influence of time-delay mismatch between the positive and negative tap groups. As shown in **Fig. 10(c)**, the time-delay mismatch and the associated phase errors do not significantly affect the synthesis of the Fano amplitude response. Instead, they mainly degrade the ER and hence the SR and ROR. Specifically, a time-delay mismatch of ~10 ps would reduce the ER from ~25.9 to ~23.8 dB when $f_c$ = 5 GHz, close to the measured value in **Fig. 7(b)**. Since other non-ideal factors may also contribute to the ER degradation, the actual time-delay mismatch is expected to be smaller, likely at the picosecond level. The time-delay mismatch can be mitigated by introducing a tunable optical delay line [54] for more precise delay control, which will be the subject of our future work.

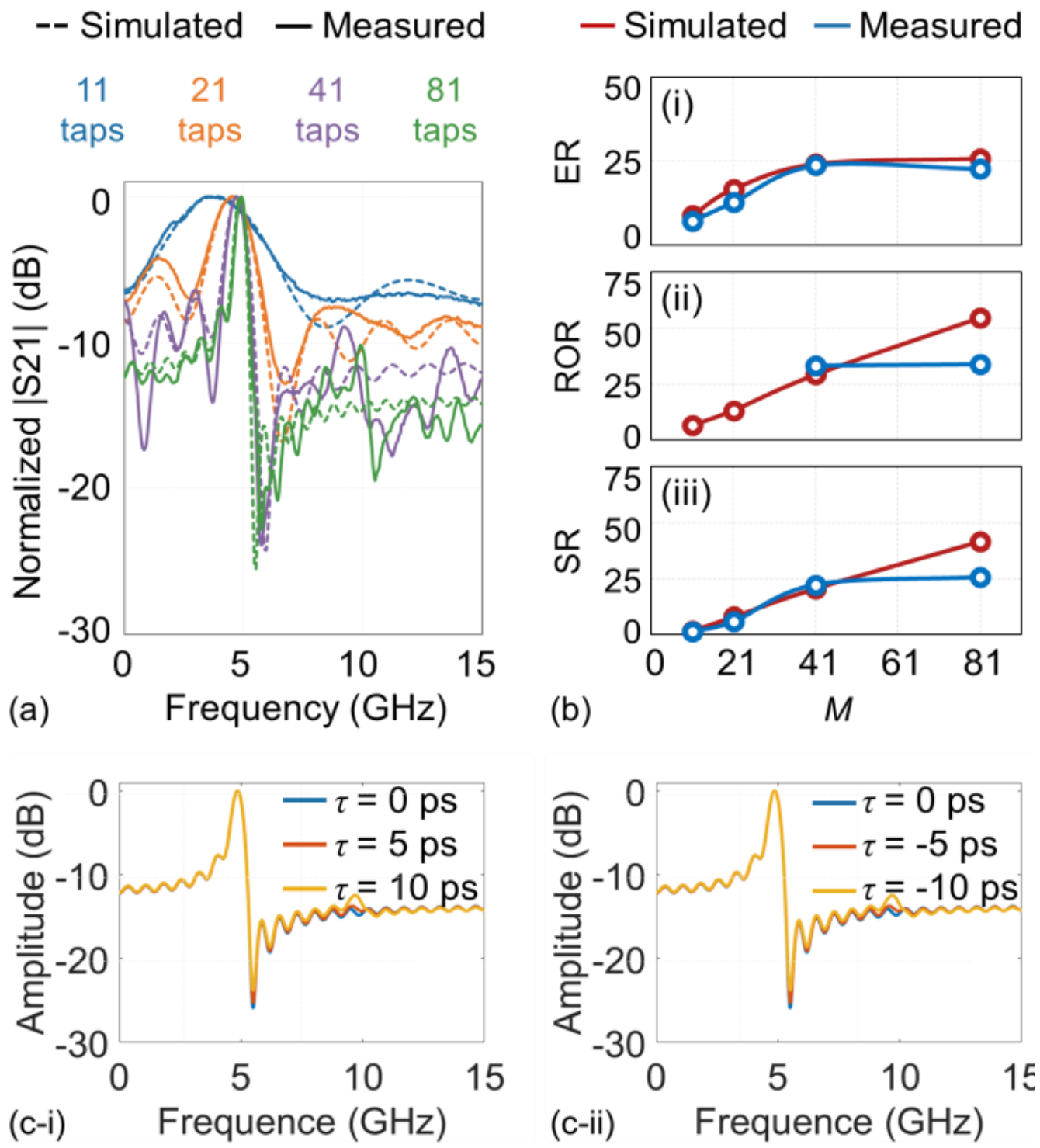

**Fig. 10.** (a) Comparison between the simulated Fano filter response in **Fig. 3(b)** and experimentally measured response in **Fig. 7(b)**, for tap numbers of $M$ = 11, 21, 41, and 81. (b) Extracted (i) ER, (ii) ROR, and (iii) SR versus $M$. (c) Simulated response of microcomb-based MWP transversal filters under different time-delay mismatches between the positive and negative tap groups, where (i) and (ii) show the results for $\tau$ = 0 ps, 5 ps, 10 ps and $\tau$ = 0 ps, -5 ps, -10 ps, respectively, with $\tau$ denoting the relative delay of the negative tap groups with respect to the positive tap groups.

IFM is a representative application of MWP Fano filters. Generally, MWP IFM systems can be implemented using either phase-modulation-based [9] or intensity-modulation-based [32] approaches. For the former, a frequency-dependent optical discriminator converts phase modulation into intensity variation for frequency estimation. This avoids bias drifts of intensity modulators and can achieve high sensitivity with a steep discriminator response, but typically requires well-defined phase response and precise resonance alignment. Whereas for the latter, the MW frequency is identified through direct amplitude-to-frequency mapping, which features a simple architecture but is susceptible to optical and MW power fluctuations.

The microcomb-based MWP Fano filters investigated in this work are more suitable for intensity-modulation-based IFM. By introducing a flat reference channel, an amplitude comparison function (ACF) can be constructed from the power ratio between the Fano-filtered and reference outputs, thereby mitigating the effects of MW input power variations. Compared with MWP Fano filters based on direct mapping the response of IIR optical filters, the FIR-synthesized Fano response can provide a high degree of flexibility by changing the center frequency (simply through programing the spectral shaper) and a high stability for long-term operation (due to inherently avoiding the need for precise wavelength alignment). These features, together with the high ROR and SR demonstrated in our experiments, enable our Fano filters to achieve high frequency measurement resolution, a large frequency measurement range, and excellent operational stability.

Monolithic integration of the microcomb-based signal processing systems represents a promising field that has witnessed substantial progress in recent years [55, 56], and developing microcomb-based MWP transversal filter systems toward a higher integration level is an important direction of our future work. On-chip integration of the dispersive delay elements can be realized using spiral waveguide arrays [55] and chirped Bragg gratings [57]. Integrated EOM can be implemented using silicon or lithium niobate modulators [58]. Programmable spectral shaping based on MRR arrays [59] and integrated photodetectors [60] have also been demonstrated. Together, these advances pave the way towards the realization of monolithically integrated microcomb-based MWP Fano filters.

Finally, it is worth noting that, current implementations of on-chip microcomb-based transversal filter systems [61] still face practical challenges in fully realizing their potential, such as limited numbers of available taps, precise wavelength alignment for on-chip spectral shaping modules, and hybrid integrations of modules made from different materials. In this context, although the discrete-component system in this work is relatively bulky and power-intensive, it can provide several advantages for practical applications at the current stage, including a large number of available taps that allow the full potential of the transversal filter architecture to be exploited, operation without the need for precise wavelength alignment during spectral shaping, convenient implementation of feedback control for system-level optimization, and high stability in long-term operation.

## VI. Conclusion

In summary, we propose and experimentally demonstrate highly reconfigurable MWP Fano filters with steep spectral transitions based on a microcomb-driven transversal filter system. Benefitting from the large number of comb lines provided by optical microcombs, the transversal filter system can synthesize filter response that closely resembles Fano resonances. In addition, the filter response is highly reconfigurable by simply programming the tap coefficients without changing any hardware. In our experiments, we demonstrate steep spectral transitions for our MWP Fano filters, achieving high roll-off rates and slope rates up to ~33.8 dB / GHz and ~25.7 dB / GHz, respectively. We also demonstrate a high degree of reconfigurability for the filter response, achieving independent tuning of all three Fano characteristic parameters including the asymmetry factor, resonance linewidth, and center frequency. The MWP Fano filters in this work provide a new route towards realizing highly reconfigurable MWP Fano filters with steep spectral transitions, which are versatile for addressing varied requirements in practical applications.

**Qi Zou** received his bachelor's degree from the Wuchang Shouyi University, China in June 2019, and his master's degree from the Shenzhen University, China in June 2023. He is currently a Ph.D. candidate at the Optical Sciences Centre in Swinburne University of Technology under the supervision of Prof. David J. Moss and Dr. Jiayang Wu. His current research focuses on optical microcombs, microwave photonics, and neuromorphic computing.

**Jiayang Wu** (Senior Member, IEEE) received the B.Eng. degree in September 2010 from Xidian University, Xi'an, China, and the Ph.D. degree in December 2015 from Shanghai Jiao Tong University, Shanghai, China. After that, he joined the Swinburne University of Technology and became a Postdoctoral Research Fellow in 2016. He is currently a Senior Research Fellow and Senior Lecturer at the Optical Sciences Centre of Swinburne University of Technology. His current research fields include integrated photonics, 2D materials, and nonlinear optics. As of January 1st of 2026, he has 108 publications in SCI journals, highlighted by *Nature*, *Nature Reviews Chemistry*, *Advanced Materials*, *Advances in Optics and Photonics*, *Nature Communications*, *Applied Physics Reviews*, *Light: Advanced Manufacturing*, *Nano Letters*, *Small*, and *Laser & Photonics Review*. He is also an inventor on 15 filed technological patents. In 2021, Dr. Wu was awarded the Australian National Research Award as the top researcher in the field of Optics & Photonics (only 1 person). From 2021 to 2025, He was consecutively named in the world's top 2% of scientists list (*Stanford University & Elsevier science-wide author databases of standardized citation indicators*).

**Yang Sun** received her bachelor's degree from the Beijing Institute of Technology, China in 2016, her master's degree from the Beijing University of Posts and Telecommunications, China in 2019, and her Ph.D. degree from Swinburne University of Technology in 2024. She is currently a postdoctoral research fellow at the Optical Sciences Centre in Swinburne University of Technology. Her current research focuses on optical microcombs, microwave photonics, and neuromorphic computing.

**Yang Li** received his bachelor's degree from the University of Jinan, China in 2015, his master's degree from the University of New South Wales, Australia in 2021, and his Ph.D. degree from Swinburne University of Technology in 2025. He is currently a postdoctoral research fellow at the Optical Sciences Centre in Swinburne University of Technology. His current research focuses on optical microcombs, microwave photonics, and neuromorphic computing.

**Guanghui Ren** received the Ph.D. degree in 2016 from RMIT University, Melbourne, VIC, Australia, where he is currently working as a Senior Research Fellow in the Integrated Photonics and Applications Centre (InPAC). His research interests include integrated optics, silicon photonics, thin-film lithium niobate, bio-photonics and hybrid integration of functional materials on integrated optics platform.

**Thach G. Nguyen** biography is not available at the time of submission.

**Xingyuan Xu** is currently a professor with Beijing University of Posts and Telecommunications. He received his Ph.D. from Swinburne University of Technology. His research focuses on neuromorphic optics, optical signal processing, and optical frequency combs. He was awarded 2019 IEEE Photonics Society Graduate Student Scholarship, 2020 Iain Wallace Research Medal from Swinburne University of Technology, 2020 Extraordinary Potential Prize of Chinese Government Award for Outstanding Self-financed Students Abroad, and 2021 *IEEE Journal of Lightwave Technology* best paper award.

**Bill Corcoran** biography is not available at the time of submission.

**Sai Tak Chu** biography is not available at the time of submission.

**Roberto Morandotti** (Fellow, IEEE) received the M.Sc. degree in physics from the University of Genova, Genova, Italy, in 1993, and the Ph.D. degree from the University of Glasgow, Glasgow, UK, in 1999. From 1999 to 2001, he was with the Weizmann Institute of Science, Rehovot, Israel. From 2002 to 2003, he was with the University of Toronto, ON, Canada, where he was involved in the characterization of novel integrated optical structures. In June 2003, he joined the Institut National de la Recherche Scientifique, Centre Énergie Matériaux Télécommunications, Varennes, QC, Canada, where he has been a Full Professor since 2008. He is the author and coauthor of more than 700 papers published in international scientific journals and conference proceedings. His current research interests include the linear and nonlinear properties of various structures for integrated and quantum optics, as well as nonlinear optics at unusual geometries and wavelengths, including terahertz. He is a Fellow of IEEE, of the Royal Society of Canada, of the American Physical Society, of the Optica (formerly the OSA), of the SPIE, and an E.W.R Steacie Memorial Fellow. He served as a Chair and Technical Committee Member for several Optica, IEEE, and SPIE sponsored meetings.

**Arnan Mitchell** was awarded the Ph.D. in engineering in 2000 from RMIT University. He is the leader of the Microplatforms research group and Director of the ARC Centre of Excellence for optical microcombs for breakthrough science (COMBS). He was Chief Investigator and RMIT Node Director of the ARC Centre of Excellence for ultrahigh bandwidth devices for optical systems (CUDOS). His research is highly multidisciplinary, spanning microfluidics, integrated optics, photonic signal processing, functional materials, microsystems, nanomaterials, and lab-on-a-chip technology. He and his team focus on platforms that enable fundamental scientific and biomedical breakthroughs producing over 300 publications. He is also committed to providing a strategic bridge between fundamental science and industry, he holds several patents and is engaged in a number of active industry projects in the fields biomedical diagnostics, communications and defence. He is the Director of RMIT's $50M MicroNanoResearch Facility which provides capabilities in integrated photonics, microfluidics, flexible electronics, ceramic microsystems, biomedical micro devices, sensors and nanoelectronics.

**David J. Moss** (Life Fellow, IEEE) is Director of the Optical Sciences Centre at Swinburne University of Technology in Melbourne, Australia, and Deputy Director of the ARC Centre of Excellence for optical microcombs for breakthrough science (COMBS). He was with RMIT University in Melbourne from 2014 to 2016, the University of Sydney from 2004 to 2014, and JDSUniphase in Ottawa Canada from 1998 to 2003. From 1994 to 1998 he was with the Optical Fiber Technology Centre at Sydney University, from 1992 ro 1994 with Hitachi Central Research Laboratories in Tokyo, Japan, and from 1988 to 1992 at the National Research Council of Canada in Ottawa. He received his Ph.D. from the University of Toronto and B.Sc. from the University of Waterloo. He won the 2011 Australian Museum Eureka Science Prize and Google Australia Prize for Innovation in Computer Science. He is a Life Fellow of the IEEE Photonics Society, Fellow of Optica (formerly the OSA), and Fellow of the SPIE. His research interests include optical microcombs, integrated nonlinear optics, quantum optics, microwave photonics, ONNs, optical networks and transmission, 2D materials for nonlinear optics, optical signal processing, nanophotonics, and biomedical photonics for cancer diagnosis and therapy.